\documentclass[letterpaper,english,aps,nofootinbib, pr, twocolumn, lengthcheck, superscriptaddress]{revtex4-2}
\usepackage{graphicx,subfigure,amsmath,amsfonts,amssymb,multirow,extarrows,bm}
\usepackage[colorlinks,linkcolor=blue,citecolor=blue,urlcolor=blue ]{hyperref}
\usepackage{bm}
\usepackage{ulem}

\begin{document}
\newcommand{\PMO}{Key Laboratory of Dark Matter and Space Astronomy, Purple Mountain Observatory, Chinese Academy of Sciences, Nanjing, 210033, People's Republic of China.}
\newcommand{\USTC}{School of Astronomy and Space Science, University of Science and Technology of China, Hefei, Anhui 230026, People's Republic of China.}
\newcommand{\RIEKN}{RIKEN Interdisciplinary Theoretical and Mathematical Sciences Program (iTHEMS), RIKEN, Wako 351-0198, Japan.}

\title{Plausible presence of new state in neutron stars with masses above $0.98M_{\rm TOV}$}
\author{Ming-Zhe Han}
\affiliation{\PMO}
\affiliation{\USTC}

\author{Yong-Jia Huang}
\affiliation{\PMO}
\affiliation{\USTC}
\affiliation{\RIEKN}

\author{Shao-Peng Tang}
\affiliation{\PMO}

\author{Yi-Zhong Fan}
\email{Corresponding author: yzfan@pmo.ac.cn}
\affiliation{\PMO}
\affiliation{\USTC}
\date{\today}

\newcommand{\redflag}[1]{{\color{red} #1}}
\newcommand{\blueflag}[1]{{\color{blue} #1}}
\newcommand{\magflag}[1]{{\color{magenta} #1}}
\newcommand{\greenflag}[1]{{\color{green} #1}}

\begin{abstract}
We investigate the neutron star (NS) equation of state (EOS) by incorporating multi-messenger data of GW170817, PSR J0030+0451, PSR J0740+6620, and state-of-the-art theoretical progresses, including the information from chiral effective field theory ($\chi$EFT) and perturbative quantum chromodynamics (pQCD) calculation.
Taking advantage of the various structures sampling by a single-layer feed-forward neural network model embedded in the Bayesian nonparametric inference, the structure of NS matter's sound speed $c_{\rm s}$ is explored in a model-agnostic way. 
It is found that a peak structure is common in the $c_{\rm s}^2$ posterior, 
locating at $2.4-4.8\rho_{\rm sat}$ (nuclear saturation density) 
and $c_{\rm s}^2$ exceeds ${c^{2}}/{3}$ at 90\% credibility. 
The non-monotonic behavior suggests evidence of the state deviating from hadronic matter inside the very massive NSs.
Assuming the new/exotic state is featured as it is softer than typical hadronic models or even with hyperons, we find that a sizable ($\geq 10^{-3}M_\odot$) exotic core, likely made of quark matter, is plausible for the NS with a gravitational mass above about $0.98M_{\rm TOV}$, where $M_{\rm TOV}$ represents the maximum gravitational mass of a non-rotating cold NS. The inferred $M_{\rm TOV} = 2.18^{+0.27}_{-0.13}M_\odot$ (90\% credibility) is well consistent with the value of $2.17^{+0.15}_{-0.12}M_\odot$ estimated independently with GW170817/GRB 170817A/AT2017gfo assuming a temporary supramassive NS remnant formed after the merger. PSR J0740+6620, the most massive NS detected so far, may host an exotic core with a probability of $\approx 0.36$.
\end{abstract}

\maketitle

\textbf{keywords:} Neutron Star; Equation of State; Quark Matter; Gravitational Wave; Bayesian Inference

\section{Introduction}

The state of strongly interacting matter at exceedingly high density remains one of the long-standing open questions.
Neutron star (NS), as it cools down the eons ahead after the birth in the supernova explosion, provides an astrophysical laboratory to investigate the equation of state (EOS) of dense, strongly interacting nuclear matter at zero temperature \citep{2017RvMP...89a5007O,2018RPPh...81e6902B,2021ARNPS..71..433L}. 
In the past five years, there has been some inspiring progress in astrophysical observations on NSs, including the multi-messenger observations of the first binary neutron star merger event GW170817 \cite{2017PhRvL.119p1101A,2017ApJ...848L..12A,2018PhRvL.121p1101A,2019PhRvX...9a1001A}, the accurate mass determination of the very massive object PSR J0740+6620 (i.e., $M=2.08 \pm 0.07 M_{\odot}$ \citep{2020NatAs...4...72C}), and the mass-radius measurements of PSR J0030+0451 and PSR J0740+6620 by the Neutron Star Interior Composition Explorer (NICER; \citep{2019ApJ...887L..21R,2019ApJ...887L..24M,2021ApJ...918L..27R,2021ApJ...918L..28M}. These events/objects comprise the multi-messenger NS data set for our following analysis.

On the theoretical side, state-of-the-art ab-initio calculation provides boundary conditions on the EOS for both low and high-density regimes.
Currently, calculations using the $\chi$EFT have been achieved with incredibly high precision (the Next-to-Next-to-Next-to leading order, N$^3$LO) for many-body interactions \citep{2019PhRvL.122d2501D}.
Thus, the dense matter EOS up to $1.1\,\rho_{\rm sat}$ is solidly constrained by the N$^3$LO $\chi \rm EFT$ calculations.
Though the pQCD is only valid at ultra-high density ($\gtrsim 40\,\rho_{\rm sat}$ \citep{2010PhRvD..81j5021K,2021PhRvL.127p2003G}), the high-order pQCD calculation still provides a reference to the non-perturbative effect at a lower density, with the chemical potential reaching 2.6 GeV, where its missing-higher-order truncation error in pQCD is comparable with the uncertainty from $\chi \rm EFT$ at 1.1 $\rho_{\rm sat}$ \citep{2022arXiv220411877G}.
Such a boundary constraint from pQCD can be pushed to a considerably lower density, even reachable in astrophysical NSs \citep{2022PhRvL.128t2701K}.
The information that emerged from various directions reveals that the EOSs, which follow the $\chi \rm EFT$ calculation in the low density, are required to undergo a rapid stiffening, and exceed the conformal limit ($c_{\rm s}^{2} /c^2 \leq 1/3$) \citep{2015PhRvL.114c1103B,2019PhRvL.122l2701M,2020PhRvC.101c5201J,2021PhRvD.104g4005K} to support a massive NS, where $c_{\rm s}$ is the speed of sound inside the NS, and $c$ is the speed of light in vacuum.
Subsequently, they must tend to be soft to satisfy the causality-driven constraint from pQCD \citep{2021PhRvL.127p2003G}.
Therefore, with the EOS structure determined by taking into account the observational and theoretical constraints, two key questions might be answered: how the quark-hadron transition takes place \citep{2018PhRvL.120z1103M,2019PhRvD..99j3009M,2020ApJ...899..164H,2021PhRvD.103f3026T,2021PhRvD.104f3032T,2013ApJ...764...12M,2018RPPh...81e6902B,2019ApJ...885...42B,2021PhRvD.104g4005K} and whether quark matter core exists in astrophysical NSs \citep{2004Sci...304..536L,2020NatPh..16..907A,2020PhRvD.101l3030F}.

\begin{figure}[htb]
\centering
\includegraphics[width=0.45\textwidth]{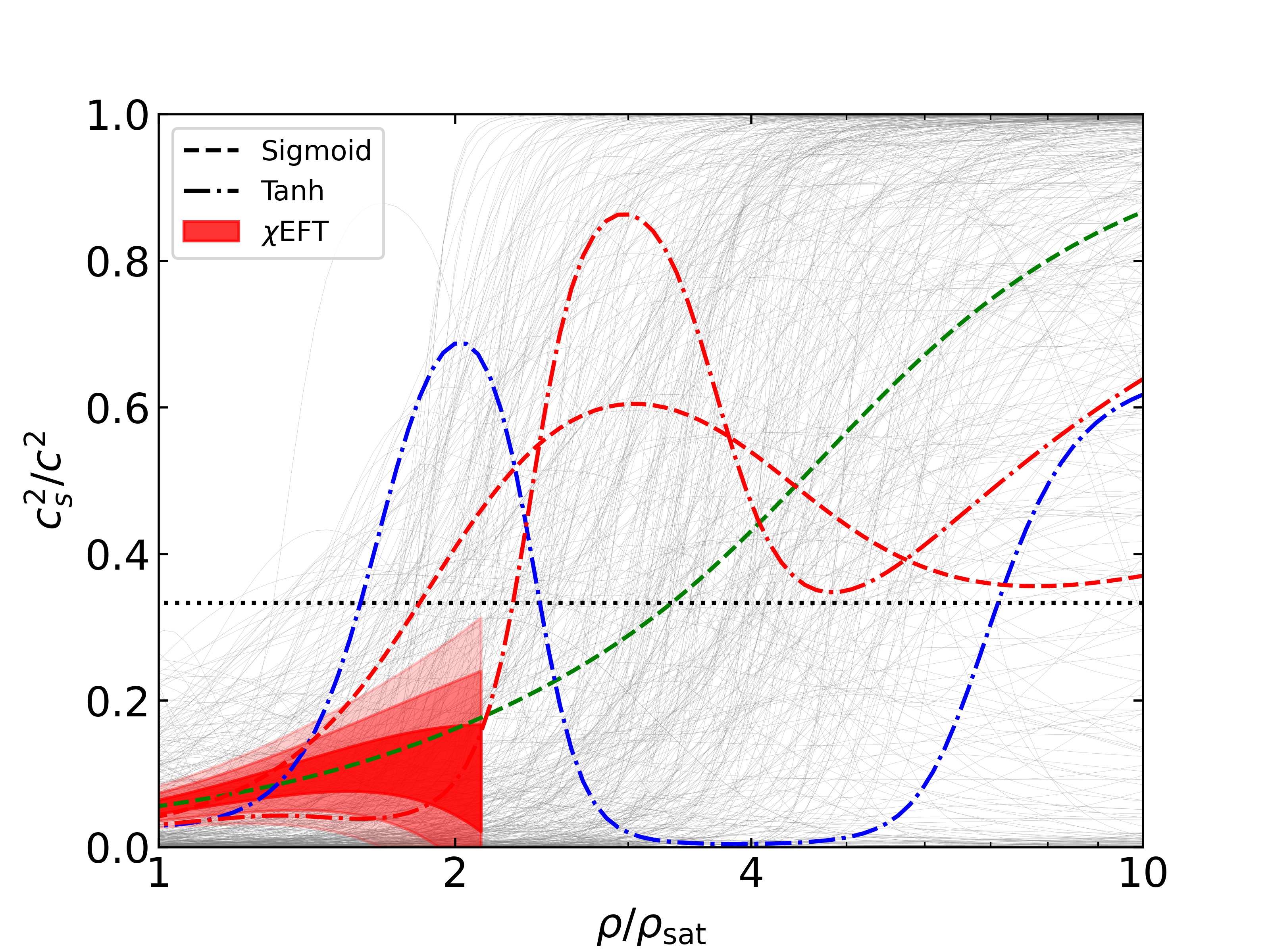}
\caption{Squared speed of sound v.s. rest-mass density for the randomly selected 500 EOSs from the prior.
Some physically-motivated EOSs, such as the hadronic (green), the first-order phase transition (blue), and the quark-hadron crossover (red) can be effectively generated in our approach.
The dashed and dash-dotted lines represent the samples generated with {\it sigmoid} and {\it hyperbolic tangents} activation functions, respectively.
The red regions with different transparencies are the  $(1,~2,~3)\sigma$ credible intervals of the $\rm \chi \rm EFT$ truncation errors \citep{2019PhRvL.122d2501D}.}
\label{fig:priors}
\end{figure}

In this work, we incorporate the latest $\chi \rm EFT$ and pQCD results/constraints in our Bayesian nonparametric inference of NS EOS represented by the feed-forward neural network (FFNN) expansion \citep{2021ApJ...919...11H} and then apply such an inference to the current multi-messenger NS data set. 
Different from the literature that assumes some structures (e.g., bumps, dips, and kinks) in the square of sound speed $c_s^2$ through a parametric form (e.g., \citep{2020PhRvL.125z1104T}), our nonparametric representation of EOS is model-agnostic and can directly/robustly extract the structure information from the observation data. We notice a peak in the reconstructed $c_{\rm s}^{2}$ curve at densities of $2-5\rho_{\rm sat}$, with $c_{\rm s}^{2}$ growing more rapidly than many pure-hadron matter models and breaking the conformal limit at $90\%$ credibility.
At density $\rho\sim5-10\rho_{\rm sat}$, the pQCD constraint \citep{2022PhRvL.128t2701K} drives $c_{\rm s}^{2}/c^2 \lesssim 0.6$. 
Motivated by Ref.~\citep{2020NatPh..16..907A} and the investigation in this work, we suggest that {\it a new/exotic state (likely the quark matter) presents when the polytropic index $\gamma\leq 1.6$ is continuously satisfied to the asymptotic densities and $c_{\rm s}^2/c^2\leq 0.7$}.
Then we show that sizable exotic cores are plausible ($\geq 90\%$ probability) for the NSs heavier than $0.98M_{\rm TOV}$, where $M_{\rm TOV}=2.18^{+0.27}_{-0.13}~M_\odot$ ($90\%$ credibility) is the maximum mass of non-rotating NSs. 
While a $\sim 10^{-3}M_\odot$ exotic core is unlikely ($\lesssim 1\%$) for the NSs lighter than $\sim 0.92M_{\rm TOV}$.

\section{Methods}

EOS provides the functional relation between the pressure $p$ and the energy density $\epsilon$, i.e., $p(\epsilon)$. Proper representations are necessary to translate the information from astrophysical data to constraints for EOSs.
So far, many phenomenological models have been proposed and can be generally divided into two categories: the parametric and nonparametric methods.
The parametric methods mainly include the piecewise polytropes \citep{2009PhRvD..79l4032R}, the spectral expansion \citep{2014PhRvD..89f4003L}, and the $c_{\rm s}^2$ based parameterizations \citep{2020PhRvL.125z1104T,2022PhRvD.105b3018T,2020NatPh..16..907A,2022ApJ...939L..34A,2022ApJ...939L..35E,2023PhRvC.107b5802M,2018ApJ...860..149T,2020Sci...370.1450D,2022Natur.606..276H}.
While the nonparametric methods involve the Gaussian process \citep{2020PhRvD.101f3007E} and the feed-forward neural network (FFNN) expansion \citep{2021ApJ...919...11H,2023CoPhC.28208547S,2022arXiv220908883S}.

Below we recall our nonparametric approach presented in \cite{2021ApJ...919...11H}, in which the $\phi$ as a function of $p$ can be described by a single-layer FFNN,
\begin{equation}
    \phi(p) = \sum_i^{\rm N} w_{2i}S(w_{\rm 1i}\log p+b_{\rm 1i})+b_{2},
\end{equation}
where $\phi=\log(c^2/c_{\rm s}^2-1)$ is the auxiliary variable to ensure the microscopical stability and causality condition $0\le c_{\rm s}^2/c^2 \le 1$, $w_{\rm 1i}$, $w_{\rm 2i}$, $b_{\rm 1i}$, and $b_2$ are weights/bias parameters of the FFNN,  the number of the nodes ${\rm N}$ is chosen to be 10, and 
\begin{equation}
    S(x) = \frac{1}{1+e^{-x}}.
\end{equation}

Now we take an advanced version that is expressed as 
\begin{equation}
    c_s^2(\rho) = c^2 S(\sum_i^{\rm N} w_{2i}\sigma(w_{1i}\log\rho+b_{1i})+b_{2}),
\end{equation}
where $\sigma(\cdot)$ is the activation function.
$S(x)$ ranges from 0 to 1, guaranteeing the microscopical stability and causality condition.
The 10\textendash{}node single\textendash{}layer FFNN is still adopted for its capability to fit almost all theoretical EOSs pretty well \citep{2021ApJ...919...11H}. 
We consider two types of activation functions, including {\it sigmoid} ($S(x)$) and {\it hyperbolic tangent} (tanh(x)) that is defined as
\begin{eqnarray}
    \textrm{tanh}(x) = \frac{e^x-e^{-x}}{e^x+e^{-x}}.
\end{eqnarray}
The model with $S(x)$ is easier to mimic the monotonically increasing sound speed or with a gentle peak \citep{2019ApJ...885...42B}, while the model with $\textrm{tanh}(x)$ can effectively generate the EOSs with an exotic structure like the zero sound speed or a sharp peak (see Fig.~\ref{fig:priors}). 
Therefore, a combination of these two models considerably enlarges the prior space.
The inferred NS's properties with each activation function are consistent, we thus combine the two sets of posteriors to obtain the results.
As shown in Fig.~\ref{fig:priors}, we randomly select 500 EOSs from the prior, which could fill the space in $c_{\rm s}^2$ after $\sim 2 \rho_{\rm sat}$.  
We mark four typical EOSs from {\it hyperbolic tangent} and {\it sigmoid} activation functions by dash-dotted and dashed lines, respectively. Their possible physical clarification is marked with different colors.

The EOS model is constructed on a log-uniform grid in densities between $\sim 0.3 \rho_{\rm sat}$ and $10\rho_{\rm sat}$.
We match the constructed EOS to the NS crust EOS \citep{1971ApJ...170..299B,2001A&A...380..151D}, up to $\sim 0.3\rho_{\rm sat}$.
From $0.3\rho_{\rm sat}$ to $1.1-2\rho_{\rm sat}$, we follow the N$^3$LO $\chi$EFT calculation \citep{2019PhRvL.122d2501D}. 
At higher densities, the $\chi$EFT calculations are likely broken down.
Though the result is weakly dependent on the choice of breakdown density \citep{2021ApJ...922...14P,2020PhRvC.102e5803E}, we define a variable $\rho_{\rm ceft}$ to marginalize the uncertainties, and uniformly sample it from $1.1~\rho_{\rm sat}$ to $2~\rho_{\rm sat}$.
Below $\rho_{\rm ceft}$ the EOSs are constrained by the $\chi$EFT calculations.
Hence the influence of the different breakdown densities of the $\chi$EFT calculations has been considered in this work.

Once the EOS is constructed, we can then predict the relations between the macroscopic properties of NS, which can be used to perform the Bayesian inference to obtain the posterior distributions of the EOS given the observation data.
The overall likelihood of the Bayesian inference is expressed as
\begin{eqnarray}
    \mathcal{L} &= \mathcal{L}_{\rm GW}
    \times \mathcal{L}_{\rm NICER}
    \times \mathcal{L}_{\rm \chi \rm EFT}
    \times \mathcal{L}_{\rm pQCD}.
\end{eqnarray}
This likelihood consists of the following parts:
\begin{itemize}
    \item $\mathcal{L}_{\rm GW}=\mathcal{P}(m_1, m_2, \Lambda_1(m_1, \theta_{\rm EOS}), \Lambda_2(m_1, \theta_{\rm EOS}))$ is the marginalized likelihood of the GW170817 interpolated by the random forest \citep{2020MNRAS.499.5972H}, where $m_{1, 2}$ and $\Lambda_{1, 2}$ are the mass and tidal deformability of the primary/secondary NS in GW170817, and $\theta_{\rm EOS}$ is the set of FFNN parameters, i.e., the weights and bias.
    
    \item $\mathcal{L}_{\rm NICER}=\prod_i \mathcal{P}_i(M(\theta_{\rm EOS}, h_i), R(\theta_{\rm EOS}, h_i))$ is the likelihood of the NICER observations, where $M$, $R$, and $h$ are the mass, radius, and core pseudoenthalpy \citep{2012PhRvD..86h4003L,2014PhRvD..89f4003L} of the NS, respectively.
    We use the Gaussian kernel density estimation (KDE) of the public posterior samples of the data from two observations, PSR J0030+0451 \citep{2019ApJ...887L..21R} and PSR J0740+6620 \citep{2021ApJ...918L..27R}. Since these data are consistent with Miller et al. \citep{2019ApJ...887L..24M,2021ApJ...918L..28M},  here we do not adopt the later as they would yield rather similar results, as found in \citep{2020ApJ...892...55J}.
    
    \item $\mathcal{L}_{\rm \chi \rm EFT}=\mathcal{P}(\epsilon, p, \rho_{\rm ceft}, \theta_{\rm EOS})$ is the likelihood considering the $\rm N^3LO$ calculation results of the $\chi$EFT theory.
    We use the publicly available samples provided in \href{https://github.com/buqeye/nuclear-matter-convergence}{Github} of Ref.~\citep{2020PhRvL.125t2702D} to obtain the means and standard deviations of the pressure.
    Then we define the $\mathcal{L}_{\rm \chi \rm EFT}$ as $1$ only if the constructed EOS falls into the 3 $\sigma$ interval of the pressure, otherwise as $0$.
    This implementation of likelihood would give almost the same results as using the full information provided in \citep{2020PhRvL.125t2702D}, while reducing the computational costs \citep{2022arXiv221100018J}.
    \item $\mathcal{L}_{\rm QCD}=\mathcal{P}(\rho_0, \epsilon(\rho_0, \theta_{\rm EOS}), p(\rho_0, \theta_{\rm EOS}))$ is the likelihood of implementing the pQCD constraints at $\sim 40\rho_{\rm sat}$, where $\rho_0$ is the rest\textendash{}mass density of the last point of the constructed EOS, the $\epsilon$ and $p$ are the corresponding energy density and pressure.
    As for the implementation of the likelihood, we use the public code released on \href{https://github.com/OKomoltsev/QCD-likelihood-function}{Github} \citep{2021PhRvL.127p2003G,2022arXiv220411877G,2022PhRvL.128t2701K}.
\end{itemize}

\begin{figure}[!htb]
\centering
\includegraphics[width=0.5\textwidth]{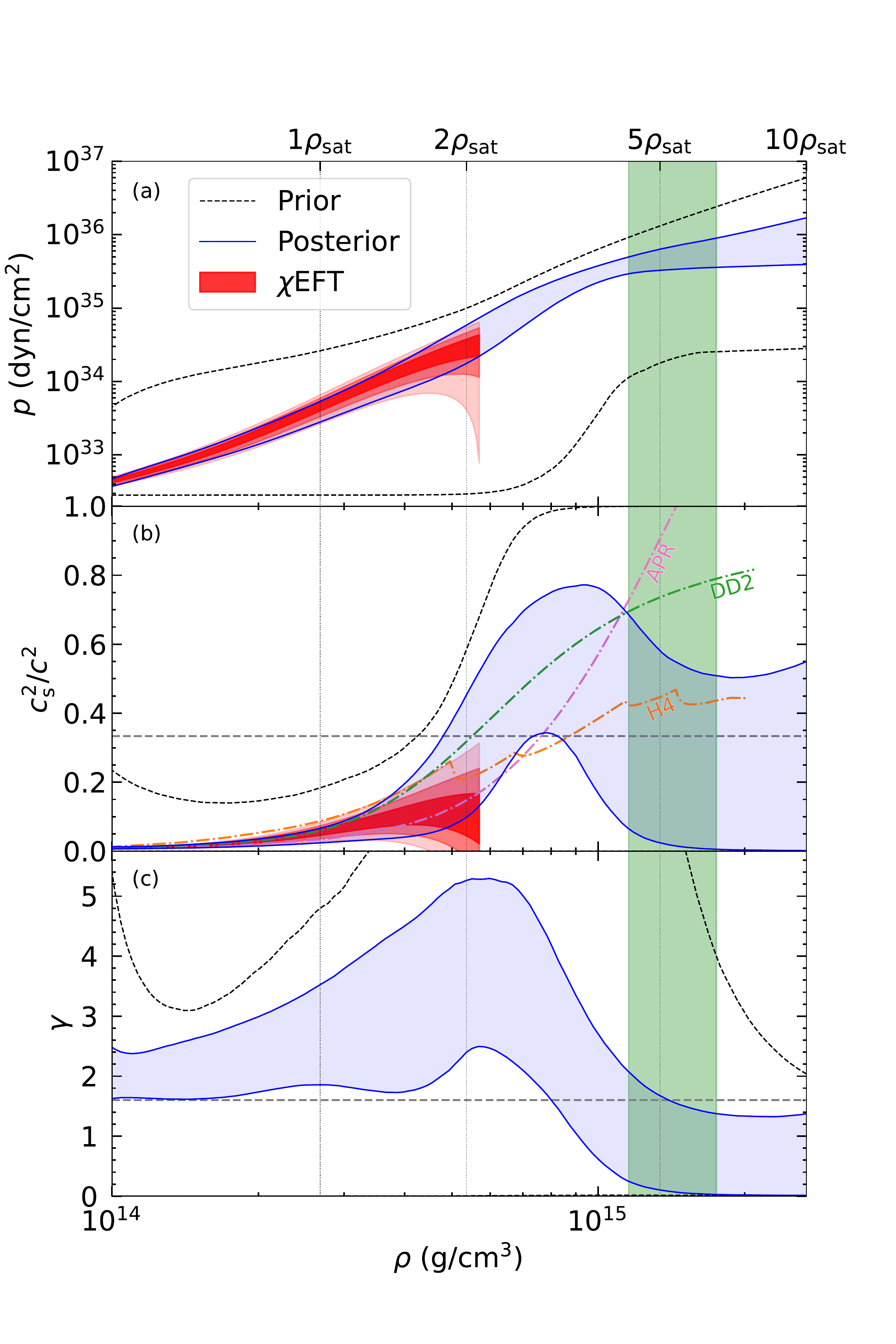}
\caption{The $90\%$ credible intervals of the pressure $p$ (panel (a)), the square of sound speed normalized by the squared light speed $c_{\rm s}^2/c^2$
(panel (b)) and $\gamma$ (panel (c), $\gamma \equiv {\rm d}(\ln p)/{\rm d}(\ln \epsilon)$, where $\epsilon$ is the energy density.) v.s. rest-mass density. 
In all panels, the vertical dotted lines mark several nuclear densities and the green vertical region denotes the central density of the heaviest NS. 
The blue regions represent the posteriors and the black dashed curves are the edge of the priors (note that in panels (b) and (c), the lower bounds of the priors are very close to zero).
The red regions with different transparencies in panels (a) and (b) are the 1, 2, and 3 $\sigma$ credible intervals of the $\rm \chi \rm EFT$ truncation errors \citep{2019PhRvL.122d2501D}.
The horizontal dashed line in panel (b) is the so-called conformal limit, i.e., $c_{\rm s}^2/c^2 \leq 1/3$, and the dash-dotted lines with different colors are the $c_{\rm s}^2/c^2$ of three representative hadronic EOSs, i.e., APR, DD2, and 
H4. In panel (c) the horizontal dashed line represents the threshold $\gamma=1.6$ of the onset of the new state. 
}
\label{fig:post}
\end{figure}

The priors of the parameters in FFNN are set as the same as Ref.~\citep{2021ApJ...919...11H}, i.e., all the parameters of the FFNN are uniformly sampled in $(-5, 5)$.
We use the Bayesian inference library {\sc BILBY} \citep{2019ApJS..241...27A} with the sampling algorithm {\sc PyMultiNest} \citep{2014A&A...564A.125B} to obtain the posterior samples of the EOSs.
The EOS with $M_{\rm TOV}$ 
beyond $1.4-3~M_\odot$ is discarded during the inference.

\section{Results and discussions}

As shown in the panel (a) of Fig.~\ref{fig:post}, the EOSs below the density of $\rho \sim 1.1\rho_{\rm sat}$ are well constrained.
This is anticipated since the $\rm \chi EFT$ theory sets a tight constraint in this range; the sound speed thus lies on the low-value region of the priors.
While in the middle region, there is a rapid increase of $c_{\rm s}^2$ and the conformal limit has been violated at the $90\%$ credibility.
As a reference, the APR \citep{1998PhRvC..58.1804A} EOS with $R_{1.4} = 11.35~\rm{km}$, DD2 \citep{2014ApJS..214...22B} EOS with $R_{1.4} = 12.90~\rm{km}$,  and H4 \citep{2006PhRvD..73b4021L} EOS with $R_{1.4} = 13.69~\rm{km}$ are also shown in panel (b) of Fig.~\ref{fig:post}, where $R_{1.4}$ denotes the radius of a $1.4 M_{\odot}$ NS (as shown in Fig.S2 we have $R_{1.4}=12.42^{+0.78}_{-1.06}$ km; for some other parameters see Fig.S3). 
The rapid stiffening in the medium density, identified in a model-agnostic way, is a natural result required by the observations of NSs with a mass of $\sim 2M_\odot$.
After that, the $c_{\rm s}^2$ are suppressed in the high-density region ($\gtrsim 4\rho_{\rm sat}$), as a consequence of the inclusion of the pQCD likelihood \citep{2021PhRvL.127p2003G}.
Therefore, the hadronic EOS with a monotonically increasing sound speed is disfavored in the high-density region. 
Specifically, the $c_{\rm s}^2\rightarrow0$ only presents near the center of the heaviest NSs, thus does not support the strong first order phase transition in low-mass NSs.

\begin{figure}[htb]
\centering
\includegraphics[width=0.5\textwidth]{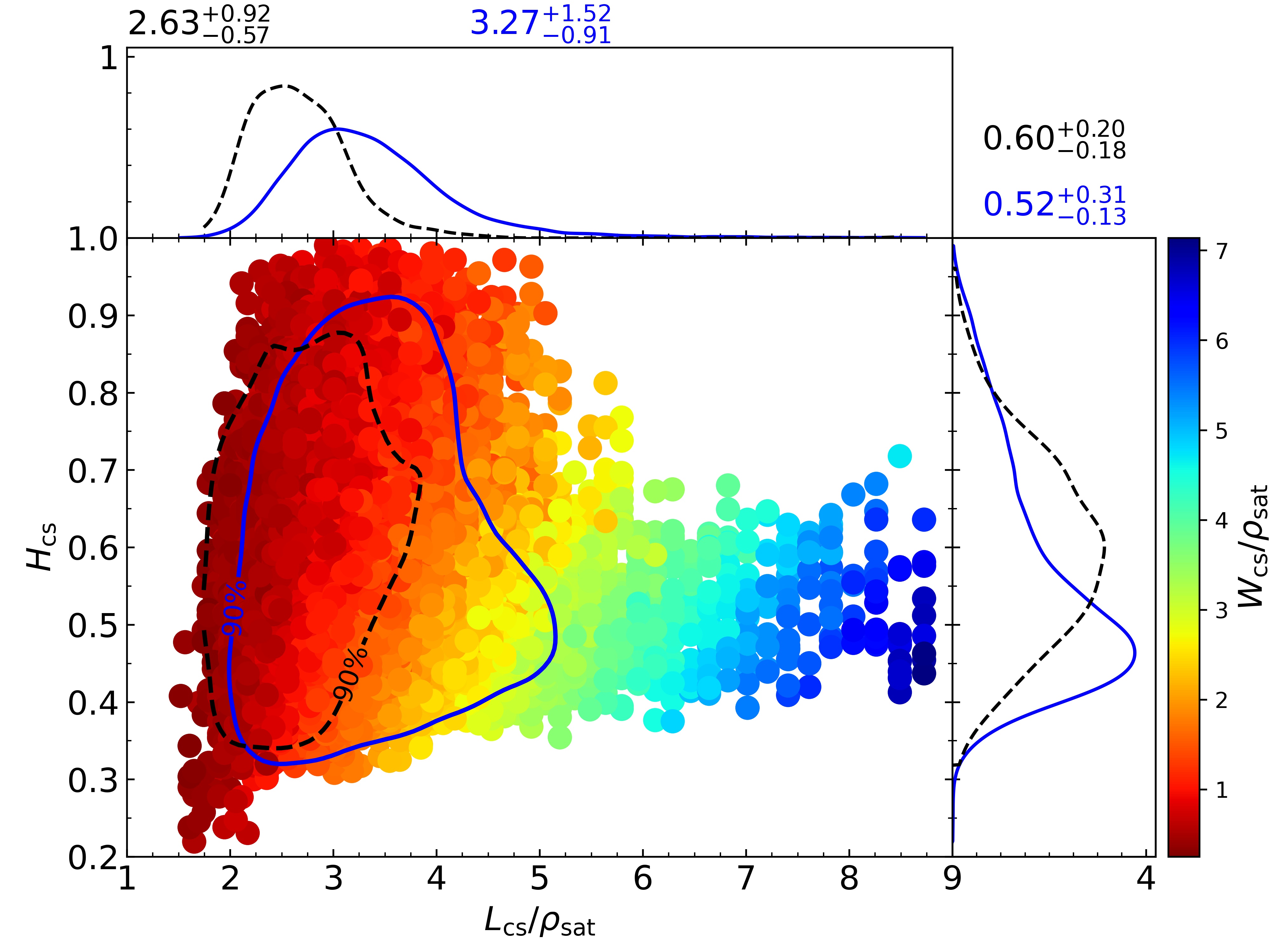}
\caption{The properties of $c_s^2$ peaks of the posterior EOSs.
$L_{\rm cs}$, $H_{\rm cs}$, and $W_{\rm cs}$ are the location, the maximum value, and the left width of the $c_s^{2}$ peak, respectively.
The $W_{\rm cs}>0$ is defined as the difference between the $L_{\rm cs}$ and the position at the half height of $H_{\rm cs}$.
The solid (dashed) line is for all posterior EOSs (the EOSs without an exotic core).
The contours correspond to $90\%$ credible regions, and so are the uncertainties of the reported values.
The one-dimensional plots are the probability density functions (PDFs).
}
\label{fig:cs_peak}
\end{figure}

The behavior of $c_{\rm s}^2$ is impacted by many factors: the $\chi \rm EFT$ in the low-density region constrains the initial condition of the sound speed, while the observed heavy NSs ($\gtrsim 2M_\odot$) and the pQCD information govern the peak's shape of the sound speed.
A peak in the $c_{\rm s}^2$ curve is quite common in our posterior EOSs.
To describe the peak structures more quantitatively, we characterize the peak of each EOS $c_{\rm s}^2$ curve with its location, height, and left width.
In particular, we define the height of peak $H_{\rm cs}$ as the maximum $c_{\rm s}^2/c^2$ of a specific EOS (only for non-monotonic EOS) and the corresponding density as the location (i.e., $L_{\rm cs}$).
We measure the rapidness of the $c_{\rm s}^2$ growth by its left width $W_{\rm cs}=L_{\rm cs}-L_{\rm hcs}$, where $L_{\rm hcs}$ represents the density when the $c_{\rm s}^2$ reaches the half of its maximum (before the peak).
The smaller the $W_{\rm cs}$ is, the more rapidly the sound speed grows.
As shown in Fig.~\ref{fig:cs_peak}, 
the $90\%$ confidence region is covered by blue contour, with $L_{\rm cs}={3.27}^{+1.52}_{-0.91}\rho_{\rm sat}$ and $H_{\rm cs}={0.53}^{+0.30}_{-0.14}$, respectively.
There is a strong positive correlation between $W_{\rm cs}$ and $L_{\rm cs}$ (also shown in Fig. S3), suggesting that $L_{\rm hcs}$ is very close to $L_{\rm cs}$ in most cases. 
The red dots do not appear for $L_{\rm cs}\geq 5\rho_{\rm sat}$, 
implying that the rapid stiffening process at a rather high density is not supported by the data.
The EOS with a late rapid stiffening process would be hard to satisfy the massive NS observation, i.e., $\gtrsim 2M_\odot$.

As discussed above, our result shows a generally more rapid growth for $c_s^2$ than that of $\chi$EFT and naturally resulting in a peak feature. 
The non-monotonic $c_s^2$ suggests the state deviating from hadronic already presents in very massive NSs. 
It is thus necessary to further examine the nature of such matter. Ref.~\citep{2020NatPh..16..907A} have discussed the possible criterion on the onset of the quark matter in the NS core. 
As shown in their Fig. 2, the polytropic indexes at the center of the most massive NS obtained in nuclear and quark matter calculations have distinct values. 
They found that the $\gamma=1.75$ is both the average between its pQCD and $\chi$EFT limits and very close to the minimal value the quantity obtains in viable hadronic models (see also their discussion in the Methods), which leads these authors to separate hadronic matter from quark matter with the criterion that $\gamma$ is continuously less than 1.75 up to asymptotic densities. 
Consequently, they concluded that the massive neutron stars are expected to host sizable quark matter cores as long as the conformal bound $c_{\rm s}^{2}\leq 1/3$ is not strongly violated.

\begin{figure}[htb]
\centering
\includegraphics[width=0.5\textwidth]{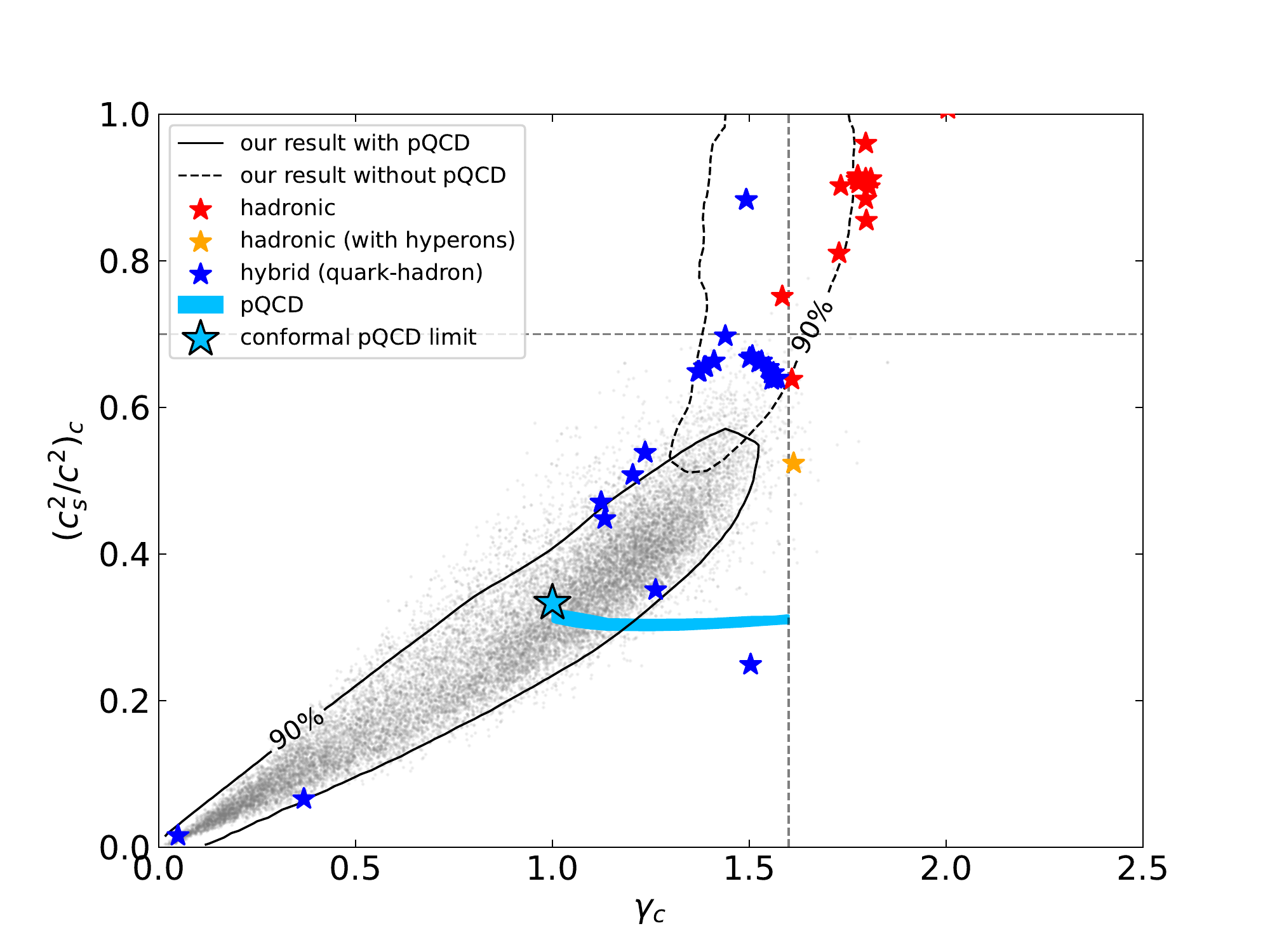}
\caption{$c_{\rm s}^{2}/c^{2}$ vs. $\gamma$ for the matter at the center of the most massive non-rotating NSs (i.e., $(c_{\rm s}^{2}/c^{2})_{\rm c}$ vs. $\gamma_{\rm c}$).
The 90\% credible regions of our results with/without pQCD likelihood are shown in black solid/dashed contour.
The red, orange, and blue stars are that calculated from theoretical hadronic \citep{2016PhRvC..94c5804F}, hadronic with hyperons \citep{2016PhRvC..94c5804F}, and hybrid \citep{2022ApJ...934...46K,2021PhRvC.103d5808D,2021PhRvD.103h6004J,2021PhRvD.103b3001B,2019ApJ...885...42B,2022PhRvX..12d1012D,2010JPhG...37i4064S,1998PThPh.100.1013S,1998NuPhA.637..435S} EOS models (these EOS tables are available at \href{https://compose.obspm.fr/home}{CompOSE}).
The high-density conformal pQCD limit is marked by a light blue star.
The vertical and horizontal dashed line denote $\gamma=1.6$ and $(c_{\rm s}^{2}/c^{2})_{\rm c} = 0.7$, respectively.
This result suggests that quark matter is the natural candidate for the exotic matter at the center of the most massive NSs, when the pQCD constraint has been incorporated.
Without the contribution of the pQCD likelihood, the nature of the core is much more uncertain.
}
\label{fig:gamma_cs}
\end{figure}

Motivated by Ref.~\citep{2020NatPh..16..907A}, in Fig.\ref{fig:gamma_cs} we show the squared speed of sound vs. $\gamma$ for the matter in the center of the most massive NSs (i.e., $(c_{\rm s}^{2}/c^{2})_{\rm c}$ and $\gamma_{\rm c}$). 
For comparison, the $(c_{\rm s}^{2}/c^{2})_{\rm c}$ and $\gamma_{\rm c}$ of some representative theoretical EOSs have also been displayed.
A group of them consists of hadrons (including hyperons), and the others are characterized by the presence of quark matter at high densities, i.e., hybrid (quark-hadron) models.
Basically we confirm the finding of Ref.~\citep{2020NatPh..16..907A} that the core of the most massive NS is hard to be made only by hadrons. 
Anyhow, we do find out that for a small fraction of hadronic EOSs, $\gamma_{\rm c}\leq 1.75$ is possible.
This was also noticed in Ref.~\citep{2020NatPh..16..907A}, and these authors argued that the constraint of $70<\Lambda_{1.4}<580$ by GW170817 \citep{2017PhRvL.119p1101A,2018PhRvL.121p1101A} had ruled out such EOS candidates.

In the current approach, for self-consistence, we take the 90\% credible ranges of $R_{1.4}$ and $\Lambda_{1.4}$ of our result, $R_{1.4}=12.42^{+0.78}_{-1.06}$ km and $\Lambda_{1.4}=467^{+223}_{-215}$, as the constraints. 
Like in Ref.~\citep{2020NatPh..16..907A}, most of the hadronic EOSs (with hyperons) with $\gamma_{\rm c}\leq 1.75$ have been excluded.
But a few, including DD2 \citep{2014ApJS..214...22B}, DDHd \citep{2004NuPhA.732...24G} and DD2Y \citep{2016PhRvC..94c5804F}, still survive because of our relatively high upper range of $\Lambda_{1.4}$ due to the inclusion of the latest NICER data.
Nevertheless, none of these EOSs falls within our 90\% credible region of the $((c_{\rm s}^{2}/c^2)_{\rm c}, ~\gamma_{\rm c})$ distribution (see Fig.\ref{fig:gamma_cs}).
As for some widely-investigated hybrid EOSs consisting of quark matter at high densities,
they are indeed characterized by $\gamma_{\rm c}\leq 1.75$, but most of them are still outside our 90\% credible region of the $((c_{\rm s}^{2}/c^2)_{\rm c}, ~\gamma_{\rm c})$ distribution.
This means that most EOSs proposed in the literature are {\it unable} to properly describe the matter at the center of the most massive NSs.
We therefore call it a ``{\it new state}", whose nature can not be uniquely determined currently from the first principle.
In view of the significant overlap of our $((c_{\rm s}^{2}/c^{2}_{\rm c},~\gamma_{\rm c})$ distribution region with that of pQCD prediction as well as some hybrid EOS models, {\it one natural speculation is that this new state is made of quark matter} (see also \cite{2020NatPh..16..907A}) though the possibility of the presence of novel interaction among the very dense matter can not be ruled out.
Indeed, the so-called conformal limit of the pQCD matter (i.e., $c_{\rm s}^2=c^{2}/3$ and $\gamma_{\rm c}=1$) is well within our favored $((c_{\rm s}^{2}/c^{2})_{\rm c},~\gamma_{\rm c})$ region.
In comparison to Ref.~\citep{2020NatPh..16..907A}, {\it now we suggest a more ``conservative" criterion for the possible onset of the exotic matter, i.e., $\gamma_{\rm c}\leq 1.6$ and $c_{\rm s}^2 \leq 0.7c^2$} to eliminate the potential contamination of a few specific hadronic EOSs (though DD2 can yield a similar $\gamma_{\rm c}$, but the predicted $c_{\rm s}^{2} \approx 0.8c^{2}$ is too high to be consistent with our bounds, see Fig.~\ref{fig:post}).

In Fig.\ref{fig:gamma_cs}, we also present the results without incorporating the pQCD constraints (i.e., the region covered by the dashed line).
Clearly, the favored $((c_{\rm s}^2/c^{2})_{\rm c},~\gamma_{\rm c})$ regions are significantly different for the scenarios with and without the contribution of the pQCD likelihood.
The soft EOSs are strongly preferred by the inclusion of the pQCD constraints.
The same conclusions can be drawn with Fig.~S4 in the Supplemental Materials, where the inclusion of the pQCD constraints yield (considerably) lower $p$, $c_{\rm s}^2/c^2$ and $\gamma$ at densities of $\rho>2\rho_{\rm sat}$ (see also  Ref.~\citep{2022arXiv220411877G} for part of these phenomena). 

\begin{figure}[htb]
\centering
\includegraphics[width=0.5\textwidth]{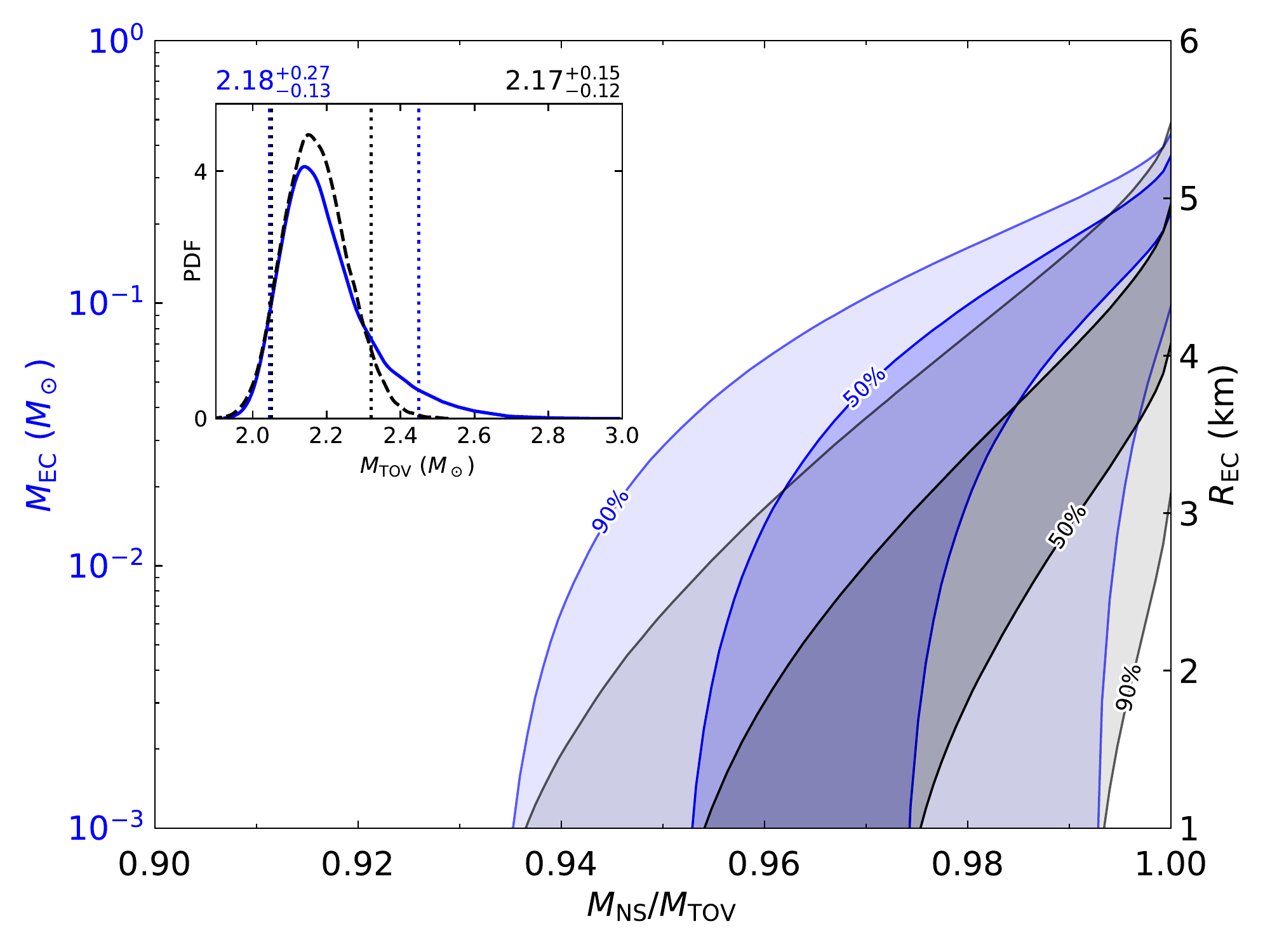}
\caption{Mass and radius of the exotic core versus $M_{\rm NS}/M_{\rm TOV}$ for a given EOS.
The blue (for mass) and black (for radius) with different transparency represent the $50\%$ and $90\%$ credible regions, respectively.
The insert presents the PDF of the inferred $M_{\rm TOV}$ (solid blue line) as well as that reported in Ref.~\citep{2020ApJ...904..119F} with an independent approach (black dashed line), and the uncertainties of the reported values are for the $90\%$ credibility. 
}
\label{fig:mq_mtov}
\end{figure}

\begin{figure}[htp!]
\centering
\includegraphics[width=0.5\textwidth]{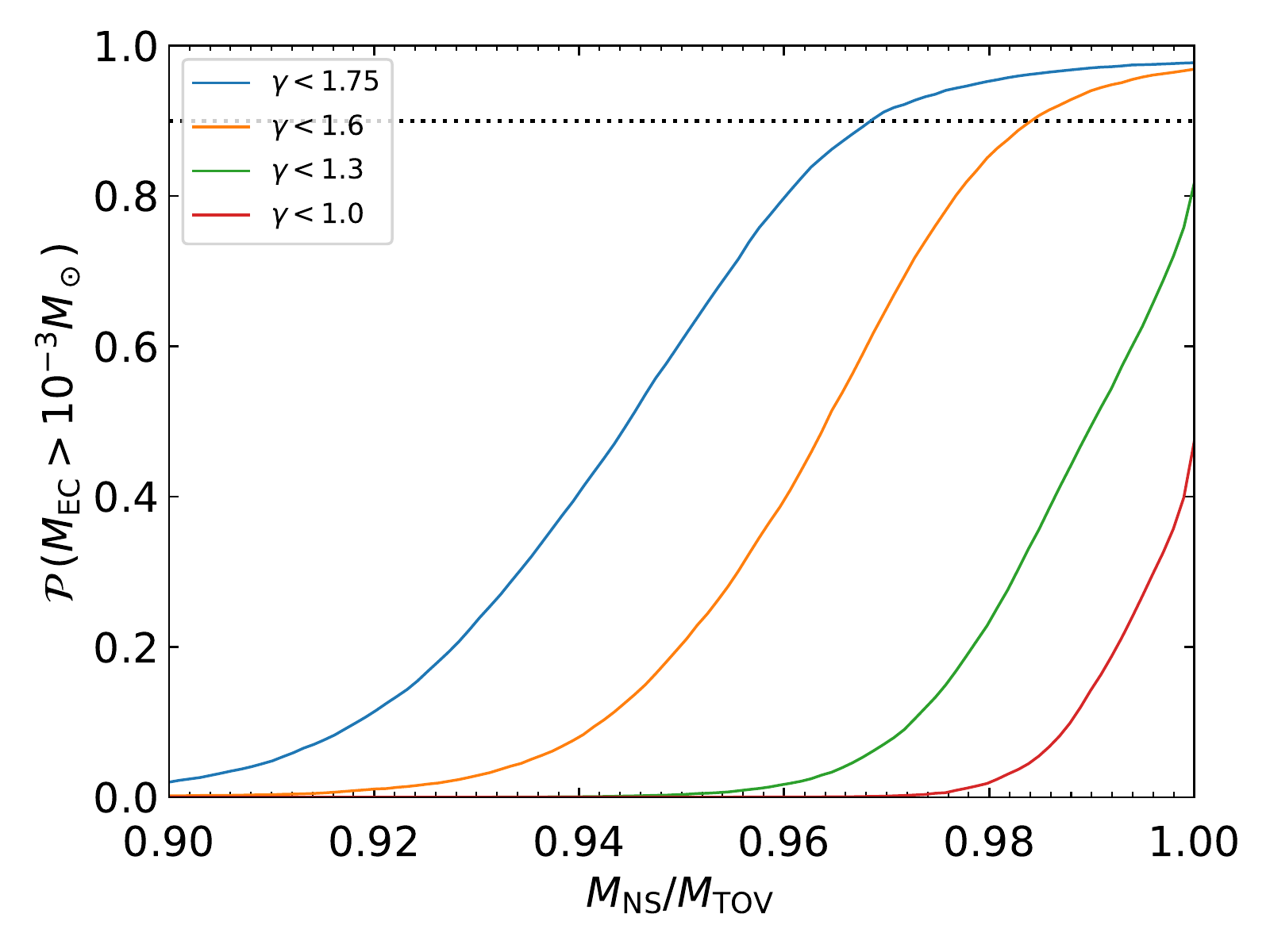}
\caption{Ratios of EOSs predicting $M_{\rm EC}>10^{-3}M_\odot$ as a function of the normalized masses $M_{\rm NS}/M_{\rm TOV}$. 
The  blue line and yellow line represent the criteria of $\gamma<1.75$ suggested in Ref.~\citep{2020NatPh..16..907A} and $\gamma<1.6$ proposed in this work, respectively. The cases of two other even more conservative criteria have been shown for illustration (see the green and pink lines). Adopting the criteria of $\gamma<1.6$ for the onset of exotic matter,  we find out that $90\%$ EOSs give $>10^{-3}M_\odot$ exotic cores for the NSs with masses of $0.984M_{\rm TOV}$.
}
\label{fig:p_quark_core}
\end{figure}

As seen from panel (c) of Fig.~\ref{fig:post}, the condition $\gamma \gtrsim 1.6$ is satisfied in density lower than $2 \rho_{\rm sat}$ (90\% credible interval), which is consistent with the systematic study for $\gamma$ of hadronic EOSs mentioned in \citep{2020NatPh..16..907A}.
Moreover, an exotic core plausibly presents inside the very massive NS because the density where $\gamma$ drops back to $1.6$ could be lower than the center density of the NS with $M=M_{\rm TOV}$.

Quantitatively, as shown in Fig.~\ref{fig:mq_mtov}, under the condition that the new state appears, the mass and the radius of the exotic core, $M_{\rm EC}$ and $R_{\rm EC}$, is sensitively dependent on the NS mass normalized by the corresponding $M_{\rm TOV}$, i.e., $M_{\rm NS}/M_{\rm TOV}$. 
The marginalized probability (see the Supplemental Materials for the detailed calculation) suggests that a $\geq 10^{-3}M_\odot$ exotic core is unlikely for $M_{\rm NS}\leq 0.92M_{\rm TOV}$,
while plausible for $M_{\rm NS}=0.984M_{\rm TOV}$ (See Figure \ref{fig:p_quark_core}).
Now the maximum mass of a non-rotating NS is constrained to be $M_{\rm TOV}={2.18}^{+0.27}_{-0.13}~M_\odot$ (90\% credibility),
which is well consistent with that inferred with the data of GW170817/GRB 170817/AT2017gfo (i.e., $2.17^{+0.15}_{-0.12}M_\odot$ in the 90\% credible interval \citep{2020ApJ...904..119F}) by assuming that the central compact collapsed into a black hole at $t\sim 0.8$ s after the merger. Our result is thus in support of the black hole central engine model for GRB 170817A (see also \citep{2022arXiv220411877G}). The consistency of $M_{\rm TOV}$ also suggests that our EOS inference of NS matter incoporating the pQCD constraints is reasonable.

\section{Conclusion}
In this work, we adopt the Bayesian nonparametric method introduced in Ref.~\citep{2021ApJ...919...11H} to constrain the EOSs and study the sound speed properties of NS matter.
We incorporate the state-of-the-art $\chi \rm EFT$ results up to $\sim 1.1-2\rho_{\rm sat}$ in the low-density range and implement the pQCD likelihood at high density.
Then, we use the mass\textendash{}tidal-deformability measurements of GW170817 and the mass\textendash{}radius of PSR J0030+0451/PSR J0740+6620 measured by NICER to perform Bayesian inference of EOS.
The sound speed properties reconstructed from the posteriors show that the maximum sound speed is above the so-called conformal limit NS at the $90\%$ credible level.
After tracking the structure of the $c_s^2$ curve for each EOS, we notice a generally more rapid growth for $c_s^2$ than that of $\chi \rm EFT$ and naturally resulting in a peak feature in most cases. 
The non-monotonic $c_s^2$ suggests the state deviating from hadronic already presents in massive neutron stars. 
Supposing the new state is featured as being softer than hadronic matter even with hyperons, we quantitatively calculate its size with the criterion $\gamma\leq 1.6$.
The results show that for $M_{\rm NS}\approx0.98M_{\rm TOV}$ a sizable exotic core, likely made of quark matter, presents at $90\%$ probability (See Figure \ref{fig:p_quark_core}).
In view of the inferred $M_{\rm TOV}=2.18^{+0.27}_{-0.13}M_\odot$ (90\% credibility), PSR J0740+6620 may have a mass exceeding such a ``threshold" and hence host an exotic core with a probability of $\approx 0.36$ (see the Supplemental Materials for the details).  Here we just use the $\chi {\rm EFT}$ result and do not take into account the heavy-ion collision data. Anyhow, the recent investigation shows that the constraints from these data are well consistent with that of the astrophysical data \cite{2022Natur.606..276H}. 
Besides $M_{\rm TOV}$, the electromagnetic radiation driven by NS mergers may also probe other aspects of NS EOS \cite{2019ApJ...877....2W}.
However it relies on some empirical relationships found in numerical simulations and suffers from the uncertainties of the reconstructed physical parameters of Gamma-ray Burst and/or kilonova ejecta.
The scientific O4 run of the LIGO/Virgo/KAGRA network \citep{2020LRR....23....3A} is upcoming this year and NICER will soon release more mass\textendash{}radius measurement results. 
Therefore, the multi-messenger data sample of NSs will increase rapidly.
Together with the new data from low-energy nuclear experiments and heavy-ion collisions experiments, more stringent constraints on the EOS will be set and the conclusions of this work will be clarified in the near future.

\section*{Acknowledgments}
The authors thank O. Komoltsev for the help in implementing the pQCD likelihood, and J.L. Jiang, T. Kojo and T. Hatsuda for the useful discussions. This work was supported in part by NSFC under grants of No. 12233011, No. 11921003 and No. 11525313.

\section*{Author contributions}
Yi-Zhong Fan, Ming-Zhe Han and Yong-Jia Huang conceived the idea. Ming-Zhe Han and Shao-Peng Tang conducted the numerical calculations. Yong-Jia Huang, Ming-Zhe Han and Yi-Zhong Fan interpreted the data. All authors discussed the results and prepared for the manuscript.

\end{document}

\newcommand{\PMO}{Key Laboratory of Dark Matter and Space Astronomy, Purple Mountain Observatory, Chinese Academy of Sciences, Nanjing, 210033, People's Republic of China.}
\newcommand{\USTC}{School of Astronomy and Space Science, University of Science and Technology of China, Hefei, Anhui 230026, People's Republic of China.}
\newcommand{\NJNU}{Department of Physics and Institute of Theoretical Physics, Nanjing Normal University, Nanjing 210046, People's Republic of China.}
\newcommand{\RIEKN}{RIKEN Interdisciplinary Theoretical and Mathematical Sciences Program (iTHEMS), RIKEN, Wako 351-0198, Japan.}
\title{Supplemental Materials}

\author{Ming-Zhe Han}
\affiliation{\PMO}
\affiliation{\USTC}

\author{Yong-Jia Huang}
\affiliation{\PMO}
\affiliation{\USTC}
\affiliation{\RIEKN}

\author{Shao-Peng Tang}
\affiliation{\PMO}

\author{Yi-Zhong Fan}
\email{Corresponding author: yzfan@pmo.ac.cn}
\affiliation{\PMO}
\affiliation{\USTC}
\date{\today}

\newcommand{\redflag}[1]{{\color{red} #1}}
\newcommand{\blueflag}[1]{{\color{blue} #1}}
\newcommand{\magflag}[1]{{\color{magenta} #1}}
\newcommand{\greenflag}[1]{{\color{green} #1}}

\maketitle
\setcounter{figure}{0}
\renewcommand{\thefigure}{S\arabic{figure}}
\renewcommand{\theequation}{S\arabic{equation}}

\section{Choice of the activation functions}
Here, we go into further detail on the techniques used in the main text.
In the main text, we extend the model developed in \citet{2021ApJ...919...11H} with some improvements. The current version of the EOS representation can be expressed as
\begin{equation}
    c_s^2(\rho) = c^2 S(\sum_i^{\rm N} w_{2i}\sigma(w_{1i}\log\rho+b_{1i})+b_{2}),
\label{eq:han_2022}
\end{equation}
where $\rho$ is the rest-mass density, $c$ is the speed of light in vacuum, $\sigma(x)$ is a nonlinear function called activation function, and $S(x)$ is the {\it sigmoid} function, 
\begin{equation}
    S(x) = \frac{1}{1+e^{-x}}.
\end{equation}

Hereafter, we define
\begin{eqnarray}
    X_i &=& w_{1i}\log\rho+b_{1i}, \\
    Y &=& \sum_i^{\rm N} w_{2i}\sigma(X_i)+b_{2}.
\end{eqnarray}
We noticed that if we take $\sigma(x) = x$, which is often called a linear activation, the functional form is very similar to the spectral expansion described in \citet{2010PhRvD..82j3011L}, i.e.,
\begin{equation}
    c_s^2(p) = c^2\left \{ 1 + \exp \left[ -\sum_k v_k \Phi_k(p) \right]  \right \}^{-1},
\end{equation}
where $\Phi_k$ is the basis function, and $v_k$ is the corresponding coefficient.
Therefore, our model is more like a general extension of the spectral expansion model.

\begin{figure}[htb]
\centering
\includegraphics[width=0.45\textwidth]{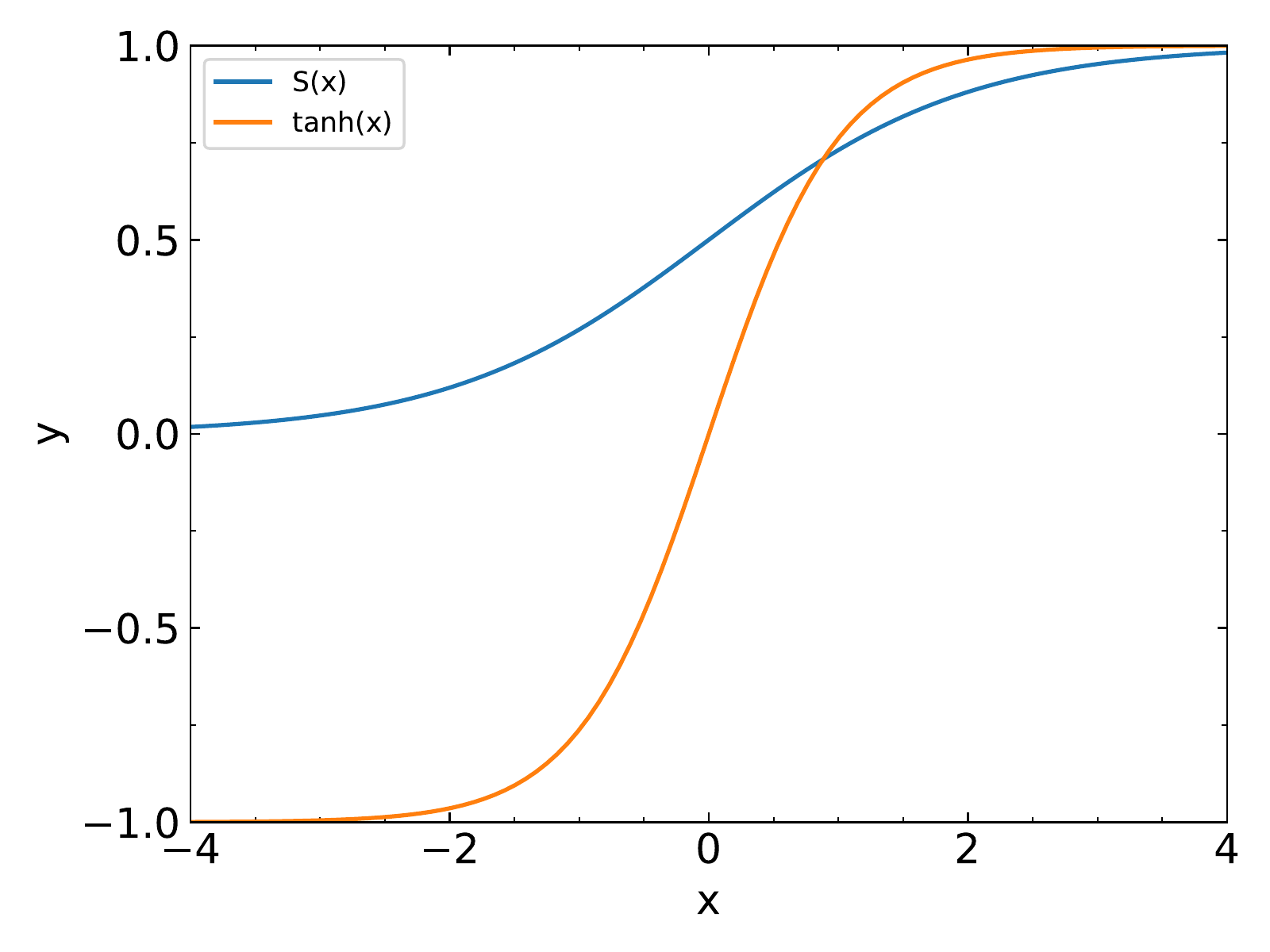}
\caption{Activation functions used in this work, i.e., {\it sigmoid} function $S(x)$ and {\it hyperbolic tangent} function $\textrm{tanh}(x)$.}
\label{fig:activations}
\end{figure}

Meanwhile, we take two kinds of activation functions in the current work, i.e., $\sigma(x)=S(x)$ and $\sigma(x)=\textrm{tanh}(x)$, where $\textrm{tanh}(x)$ is the {\it hyperbolic tangent} function,
\begin{equation}
    \textrm{tanh}(x) = \frac{e^x-e^{-x}}{e^x+e^{-x}}.
\end{equation}
These two activation functions are displayed in Fig.~\ref{fig:activations}, where we can see that both $S(x)$ and $\textrm{tanh}(x) \rightarrow 1$ when $x \rightarrow +\infty$, while when $x \rightarrow -\infty$, $S(x) \rightarrow 0$ and $\textrm{tanh}(x) \rightarrow -1$.
Given the fixed parameters of $(w_{1i}, ~w_{2i}, ~b_{1i}, ~b_2)$, i.e., when a sample of these parameters is drawn from the prior, we find that $\textrm{tanh}(X_i)$ changes more sharply than $S(X_i)$ as the independent variable (i.e., $\log\rho$) changes.
As a result, the $Y$ with $\sigma(x)=\textrm{tanh}(x)$ may be easier to get away from zero than the $Y$ with $\sigma(x)=S(x)$, which will in turn make the $c_s^2/c^2$ approaches $0$ or $1$. 
Therefore, {\it hyperbolic tangent} activation function could efficiently mimic the $c_s^2$ behaviors of the first-order phase transition ($c_s^2/c^2\sim0$), or a smooth quark-hadron crossover (QHC) with sharp peak ($c_s^2/c^2\sim 1$).
On the other hand, the model with {\it sigmoid} activation function prefers to generate the EOSs that asymptotically approach median $c_s^2$. So it could generate the pure-hadron-like EOSs (with a monotonically increasing sound speed) or a QHC with a gentle bump.
Combined with the two activation functions, the prior space would be large enough to include the general types of physically-motivated EOSs.

\section{General constraints on NS properties and EOS parameters}
\begin{figure}[htb]
\centering
\includegraphics[width=0.45\textwidth]{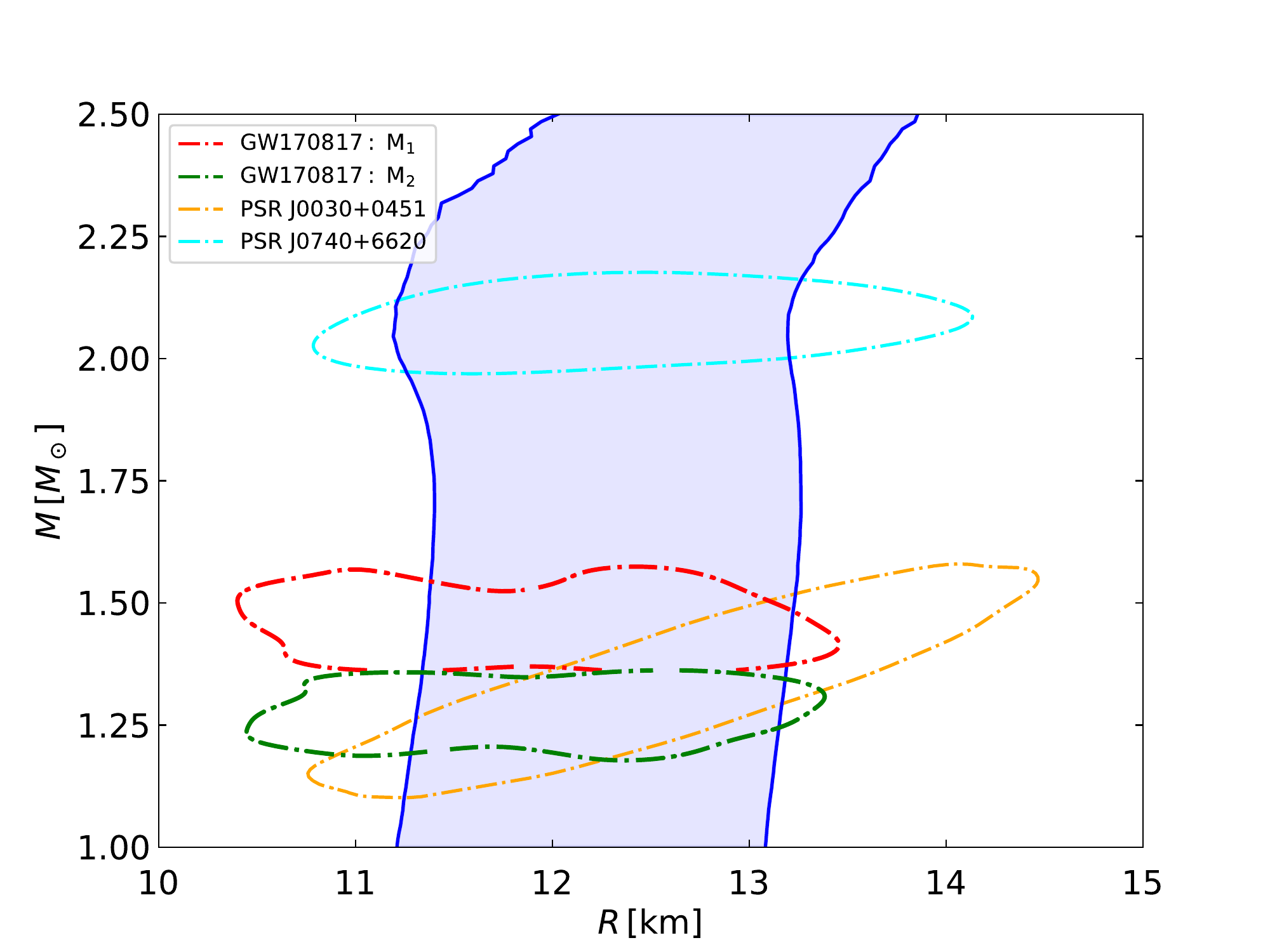}
\caption{The $90\%$ credible intervals (blue shadowed area) of radius R as a function of mass M. The dash-dotted contours ($68\%$ confidence level) in red, green, orange, and cyan are the mass\textendash{}radius measurements of the primary, the secondary object of GW170817 (data taken from the right panel of Fig.~3 of \citet{2018PhRvL.121p1101A}), PSR J0030+0451, and PSR J0740+6620, respectively.}
\label{fig:mr_range}
\end{figure}

In addition to the constraints on the EOSs mentioned in the main text, we provide more information from our model-agnostic Bayesian inference. 
In Fig.~\ref{fig:mr_range}, we show the $90\%$ credible region of the radius as a function of mass obtained from the posterior samples. 
The remarkable observation of binary neutron star merger event GW170817 \citep{2018PhRvL.121p1101A,2019PhRvX...9a1001A}, and two $M-R$ measurements of PSR J0030+0451 \citep{2019ApJ...887L..21R, 2019ApJ...887L..24M} and PSR J0740+6620 \citep{2021ApJ...918L..27R, 2021ApJ...918L..28M} are also shown. 
The joint information provides a better constraint, with $R_{1.4} = 12.42^{+0.78}_{-1.06}$ km and $R_{2.0} = 12.30^{+0.90}_{-1.11}$ km ($90\%$ credible interval).

\begin{figure*}[htb]
\centering
\includegraphics[width=0.95\textwidth]{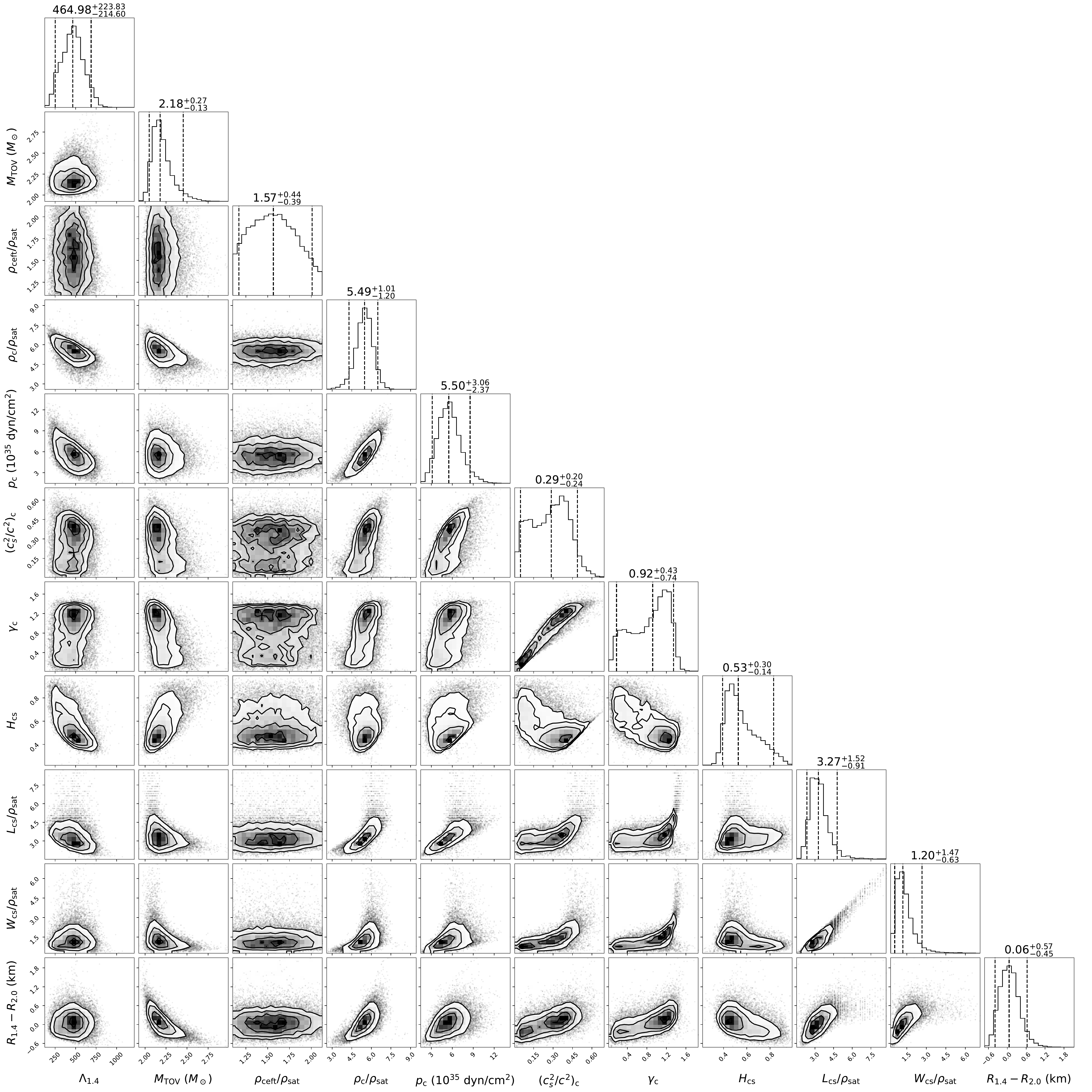}
\caption{Corner plots of the probability distributions for the tidal deformability of a canonical $1.4M_\odot$ NS ($\Lambda_{1.4}$), the maximum mass of a non-rotating NS ($M_{\rm TOV}$), the breakdown density of the $\chi$EFT calculations ($\rho_{\rm ceft}$), the rest-mass density, pressure, sound speed, and polytropic index in the center of the NS with $M=M_{\rm TOV}$ ($\rho_{\rm c}$, $p_{\rm c}$, $(c_s^2/c^2)_{\rm c}$, and $\gamma_{\rm c}$), the value of the peak in sound speed ($H_{\rm cs}$), the location of the peak in sound speed ($L_{\rm cs}/\rho_{\rm sat}$), the left width of the $c_s^2$ peak ($W_{\rm cs}/\rho_{\rm sat}$), and the difference between the radii of a $1.4M_\odot$ and a $2.0M_\odot$ NSs ($R_{1.4}-R_{2.0}$) of the EOS posterior samples.
The error bars/lines are all at the $90\%$ credible level.}
\label{fig:corner}
\end{figure*}

Some additional quantities of interest with their possible correlations are presented in Fig.~\ref{fig:corner}.
$\Lambda_{1.4}$ is the dimensionless tidal deformability of a 1.4$M_\odot$ NS, $\Lambda$ is defined as $\Lambda=(2/3)k_{2}[(c^2/G)(R/m)]^5$, where $k_{2}$ is the tidal Love number \citep{2008ApJ...677.1216H,2008PhRvD..77b1502F,2009PhRvD..80h4035D,2009PhRvD..80h4018B}.
$M_{\rm TOV}$ is the maximum gravitational mass of a non-rotating NS.
$\rho_{\rm ceft}$ is the breakdown density of the $\chi$EFT calculations.
$\rho_{\rm c}$, $p_{\rm c}$, $(c_s^2/c^2)_{\rm c}$, and $\gamma_{\rm c}$ are the rest\textendash{}mass density, pressure, squared speed of sound normalized by squared light speed, and polytropic index in the center of the NS with $M=M_{\rm TOV}$, respectively.
$H_{\rm cs}$, $L_{\rm cs}$, and $W_{\rm cs}$ are, as defined in the main text, the peak value of the squared sound speed, the location where the sound speed reaches its peak value, and the left width of the $c_s^2$ peak.
Finally, we also show a parameter that measures the difference between the radius of a $1.4~M_\odot$ NS and $2.0~M_\odot$ NS, i.e., $R_{1.4}-R_{2.0}$.

As shown in Fig.~\ref{fig:corner}, the $\rho_{\rm ceft}$ has no obvious correlations with other variables.
This means that the results do not depend on the choice of the breakdown density $\rho_{\rm ceft}$, it's consistent with previous works \citep{2021ApJ...922...14P,2020PhRvC.102e5803E}.

\section{The results with/without pQCD constraint}

\begin{figure}[htb]
\centering
\includegraphics[width=0.5\textwidth]{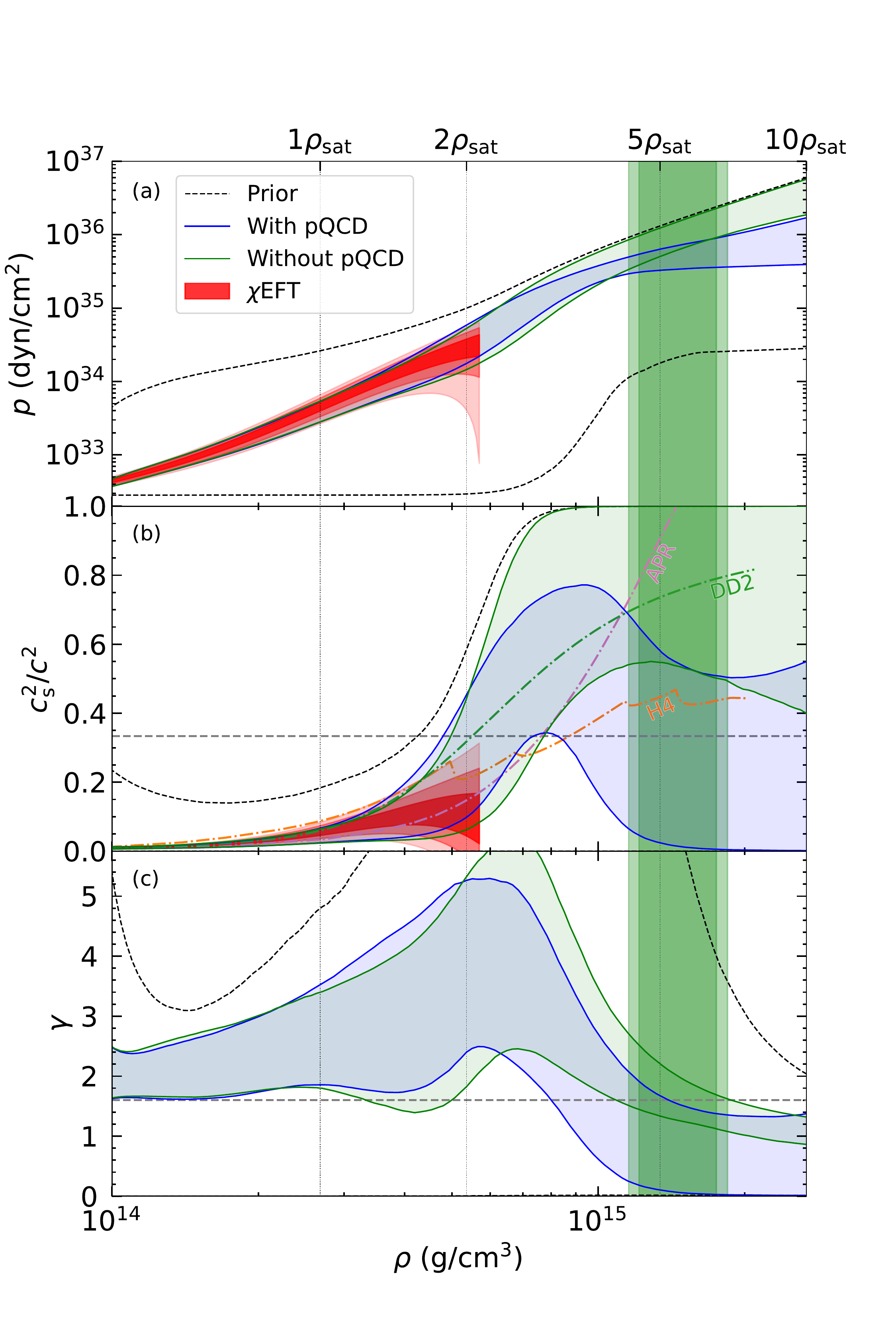}
\caption{The same as Fig.2 in the main text except including also the constraint results without incorporating the pQCD information (i.e., the light green regions). Clearly, the inclusion of the pQCD constraints yield (considerably) lower $p$, $c_{\rm s}^2/c^2$ and $\gamma$ at densities of $\rho>2\rho_{\rm sat}$, i.e., the EOSs got significantly softened.}
\label{fig:post-2}
\end{figure}

As shown in Fig.4 of the main text, the soft EOSs are strongly preferred by the inclusion of the pQCD constraints. The effects of pQCD on softening the EOSs are also evident in Fig.\ref{fig:post-2}, where the results with/without incorporating the pQCD constraints are marked in blue and light green, respectively.
Obviously, the inclusion of the pQCD constraints yield substantially lower $p$, $c_{\rm s}^2/c^2$ and $\gamma$ at densities of $\rho>2\rho_{\rm sat}$. Part of these phenomena has also been reported in Ref.~\citep{2022arXiv220411877G}.

\section{A universal relation of the exotic core?}
\begin{figure}[htb]
\centering
\includegraphics[width=0.5\textwidth]{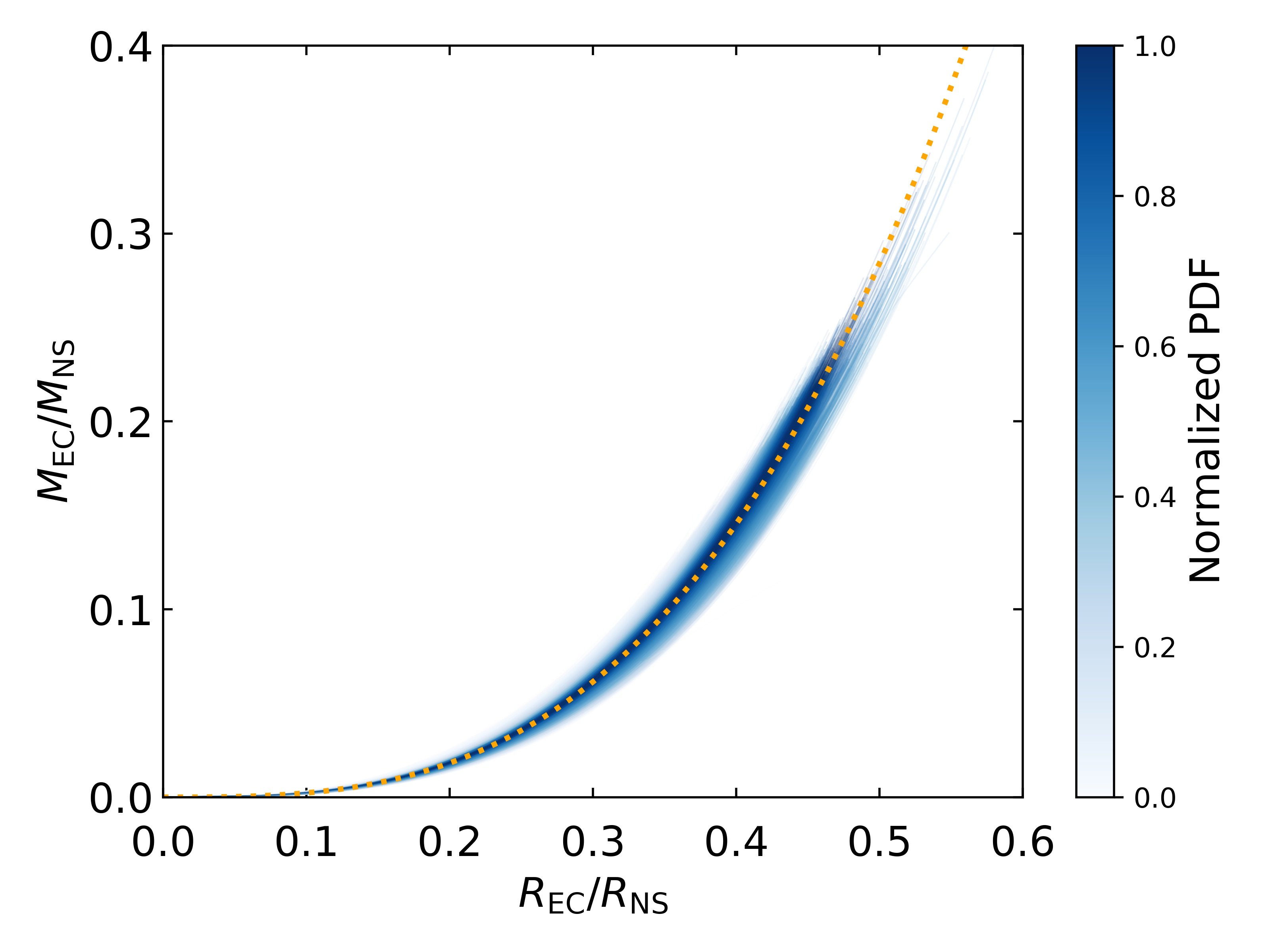}
\caption{The relation between mass and radius of the exotic core normalized to those of the host NS.
Each curve corresponds to an EOS from posteriors and is colored by its normalized PDF. The yellow dashed line represents the fit of $M_{\rm EC}/M_{\rm NS} = 2.27(R_{\rm EC}/R_{\rm NS})^3$.
}
\label{fig:quark_mr}
\end{figure}

In the presence of an exotic core, we normalize the mass and radius to that of the host NS (i.e., $M_{\rm EC}/M_{\rm NS}$ and $R_{\rm EC}/R_{\rm NS}$ and find an interesting ``universal" relation (as shown in 
Fig.~\ref{fig:quark_mr}, most of the curves are concentrated in a narrow region).
We fit the data with the function of $M_{\rm EC}/M_{\rm NS} = k(R_{\rm EC}/R_{\rm NS})^3$ and have $k=2.27^{+0.38}_{-0.36}$ ($90\%$ credibility), suggesting that the average density of the exotic core is $\sim 2.3$ times that of its host NS.
Nevertheless, we would like to remind that in a few percent of the posterior EOSs, there is no evidence for the new state onset (i.e., the $\gamma \leq 1.6$ criterion is not satisfied inside these heaviest NSs). Such EOSs are characterized by their smaller $L_{\rm cs}$ and somewhat larger $H_{\rm cs}$ compared to those hosting exotic cores, as shown by the dashed line in Fig.3.

\section{Probability for PSR J0740+6620 hosting a quark matter-like exotic core}

In the main text, we have concluded that a sizable exotic core is  plausible to be present in the very massive NS with a mass larger than about $0.98M_{\rm TOV}$.
Such a value is obtained as follows: given a series of NS masses normalized by the corresponding $M_{\rm TOV}$ of each EOS, we calculate the ratio of EOSs that predict $M_{\rm EC}>10^{-3}M_\odot$ for each value of $M_{\rm NS}/M_{\rm TOV}$ (as shown in Fig.6 of the main text); then we can directly find the percentile where almost $>90\%$ EOSs support that the NSs have exotic cores larger than $10^{-3}M_\odot$.
Based on the above results, it is pretty interesting to investigate whether (or how possible) the currently observed maximum mass NS PSR J0740+6620 hosts nonnegligible exotic matters (e.g., $>10^{-3}M_\odot$) inside its core.
The $68.3\%$ credible interval of $M_{\rm NS}/M_{\rm TOV}$ for PSR J0740+6620 is estimated to ${0.971_{-0.036}^{+0.022}}$ by reconstructing the posterior samples obtained in the Bayesian inference.
To evaluate the overall chance for PSR J0740+6620 hosting an exotic core, we marginalize the uncertainties of its mass measurement and the EOSs using 
\begin{equation}
    P=\sum_i^{N_{\rm EOS}}\mathcal{P}({\rm EOS}_{i})\frac{\int_{M_{\rm NS}(M_{\rm EC}=10^{-3}M_\odot \mid {\rm EOS}_{i})}^{M_{\rm TOV,i}}dm\,\mathcal{P}(m)}{\int_{-\infty}^{M_{\rm TOV,i}}dm\,\mathcal{P}(m)},
\end{equation}
where $\mathcal{P}(m)$ is the mass distribution of PSR J0740+6620, $M_{\rm NS}(M_{\rm EC} \mid {\rm EOS}_{i})$ is the function between $M_{\rm NS}$ and $M_{\rm EC}$ given the $i$-th EOS in the posterior samples, and $M_{\rm TOV,i}$ is the corresponding maximum mass.
Since our posterior samples are equal-weighted, $\mathcal{P}({\rm EOS}_{i})$ can be simply taken as $1/N_{\rm EOS}$, where $N_{\rm EOS}$ is the sample size.
Finally, we find that PSR J0740+6620 has a probability of $35.6\%$ for hosting a sizable exotic core if we take the criteria of $\gamma<1.6$ for the emergence of exotic matter.
Such a probability decreases to $10.0\%$ when we change the criterion to a more conservative value, say $\gamma<1.3$.

\clearpage

\bibliographystyle{apsrev4-2}
\bibliography{ref-supplemental}